\documentclass[12pt]{article}
\usepackage{amssymb}
\usepackage{amsmath,amsthm}
\usepackage[pdftex]{graphicx}
\providecommand{\U}[1]{\protect\rule{.1in}{.1in}}
\makeatletter
\def\@seccntformat#1{\csname the#1\endcsname.\quad}
\makeatother

\begin{document}

\title{{\Large \textbf{ Excitation of Flow Instabilities due to Nonlinear Scale Invariance}}}
\author{ 
Dhurjati Prasad Datta$^1$\thanks{Corresponding author; email:dp${_-}$datta@yahoo.com} \  and Sudip Sen$^{2,3}$  \\
%EndAName
 $^1$ Department of Mathematics, University of North Bengal,  \\
Siliguri, West Bengal- 734013, India \\ 
$^2$ National Institute of Aerospace (NASA-LaRC) \\
 100 Exploration Way, Hampton, VA 23666 \\
$^3$ College of William \& Mary, Williamsburg, VA 23187}
%\and \\
%Sudip Sen \\
%EndAName
%National Institute of Aerospace (NASA-LaRC), 100 Exploration Way, Hampton, VA 23666 \\
%and \\
%College of William \& Mary, Williamsburg, VA 23187
%}
\date{}
\maketitle

\begin{abstract}
A novel route to instabilities and turbulence in fluid and plasma flows is presented
in kinetic Vlasov-Maxwell model. New kind of flow
instabilities is shown to arise due to the availability  of new kinetic energy sources which 
are absent in conventional treatments. The present approach is based on a scale invariant 
nonlinear analytic formalism developed  to address  irregular motions 
on a chaotic attractor or in turbulence in a more coherent manner. We have studied two 
specific applications of this turbulence generating mechanism.
The warm plasma Langmuir wave dispersion relation is shown to become unstable in the
presence of these multifractal measures. In the second application, these
multifractal measures are shown to induce naturally non-Gaussian i.e. a
stretched -Gaussian distribution and anomalous transport for tracer
particles from the turbulent advection-diffusion transport equation in a
Vlasov plasma flow.
\end{abstract}
\begin{center}
To appear in {\bf Physics of Plasma} (2014).
\end{center}

\section{Introduction}
Relevance of intermittency and multifractal scalings in plasma fluctuations 
and turbulence have been pointed out in various recent studies. 
Data analysis based on recent satellite and spaceship measurements reveals
that space plasma fluctuations are mostly intermittent in nature with
multifractal characteristics \cite{sp1,sp2}. In a fusion  plasma
edge fluctuations in Stellarators and Tokamaks show that
the plasma turbulence is more of an intermittent nature in short
time and space scales than in moderate scales when turbulence seems to have
a monofractal feature \cite{car}. Intermittency and multifractal
fluctuations are also observed in the dynamics of discharge plasma \cite{rom}.
Significance of multifractal scalings and intermittency were studied  
in several  works in the fluid turbulence \cite%
{mandel,parisi,benzi}. The key feature of the multifractal scaling is the
nontrivial dependence of the scaling exponent $\xi (q)$  on a singularity
parameter $\epsilon$ that quantifies the strength of singularity in the multifractal measure
of the fluctuating dynamical quantity. The symbol $q$, on the other hand, denotes the 
$q$th moment for the associated probability measure and relates to the singularity parameter 
$\epsilon$ by a Legendre transform. When the exponent $\xi (q)$ scales
linearly with $q$, the fluctuation is purely self-similar, characteristic of
a monofractal behaviour. In a more general situation, $\xi (q)$ denotes  the 
generalized dimension spectrum, that is used to classify  the observed
spatio-temporal fluctuation distributions in various fluid and plasma turbulence and instabilities. 
Such fluctuation patterns do not conform to  self-similarity, but
tend, in general, to an intermittent behaviour, i.e. locally analogous to a
devil's staircase function \cite{fris}. Understanding the origin and
dynamics of such intermittent, multifractal fluctuations in the context of
fluid and plasma flows is obviously of considerable interest \cite{mandel,parisi,benzi}.

In this work we present some new analytic results opening up novel routes
to fluid and plasma instabilities \cite{plsm1} which are usually absent in a more
conventional analysis of fluid or plasma models. Our study is based on a novel 
 scale invariant nonlinear analysis developed to address in a more coherent manner the production 
of complex multifractal structures dynamically from a simple initial state. Theory of fractional kinetics \cite{klafter} using fractional calculus and fractional differential equations  are considered by various authors \cite{plsm2,klafter,chen, tar}  to offer a general theoretical framework to model anomalous scalings and
transports. The present approach is not only independent
of the fractional kinetics but is a natural extension
of the classical analysis to accommodate multifractal scaling behaviours in
a smooth (differentiable) manner (\cite{dt1} -\cite{dp3}). 
We show, in particular, how the linear differential measures of the form $dt$ or $dx_i$ in a laminar flow can be realized as smooth  multifractal measures of the form $dt^{\alpha(\epsilon)}$ and $dx_i^{\beta(\epsilon)}$, for a range of multifractal exponents $0<\alpha(\epsilon) \ , \beta(\epsilon)<1$, in a turbulent flow. The singularity parameter $\epsilon$ corresponds to the {\em scale} exposing the level of singularity that is accessed asymptotically and can be related with the Reynolds number in a turbulent flow. The derivation of such anomalous scalings rests on the assumption that increments in a complex flow could be mediated not only by linear shifts but also by small scale discrete, but {\em smooth} jumps, giving rise to a {\em duality principle} and the associated  proliferation of the underlying laminar flow differential equations into self similar replica equations over finer scales . In ordinary setting jump modes are discontinuous, asking for probabilistic arguments to extract macroscopic observables (i.e. the moments) from an irregular Brownian type flow. In the present formalism jump mediated increments are shown to satisfy simple scale invariant linear differential equation in logarithmic variables as opposed to linear shifts of the form $x\mapsto X=x+h$. Although we focus here to applications of this formalism in plasma turbulence, other potential fields of applications are evolution of biological and living systems, stock market variations, dynamics of social systems such as growth and proliferation of human habitats, internet networking, to name a few. We note in particular that the dynamics of a complex organism is known to follow an evolutionary pattern driven mainly by discrete jump mode that is  realized, however, in a smooth coherent  manner \cite{ho} avoiding direct collisions unlike the Brownian motion. The crux of the present approach is to formulate discrete small scale jumps in a smooth manner formally which would allow one to avoid direct collisions with  singular points those emerge copiously in the dynamics of a complex system. 

It is pertinent to state here the  novelty and advantages of the present approach  over those available in contemporary literature \cite{plsm2, klafter,tar,chen}. As stated already, a major aim is to develop a nonlinear differential analysis to study {\em local} behaviour of  emergent multifractal measures in turbulent media. The turbulent systems are traditionally studied by stochastic methods \cite{parisi,benzi,fris}. Mathematical models \cite{plsm2, klafter,tar,chen} based on fractional calculus, on the other hand, are known to derive anomalous multifractal scaling laws of turbulent flows directly from the underlying fractional differential equations. However, there are  various definitions of fractional derivatives such as Riemann-Liouville, Caputo, Weyl, Riezs and others, having relative merits and demerits over each other over their scope and applicability \cite{tar}. Moreover, fractional derivatives are basically non-local, thus defeating the very motivation of using a differential equation to discuss local structures. The present approach, however,  formulates a rigorous framework of using  integer-order differential equations based instead on multifractal measures for  local descriptions of evolutionary patterns of emergent complex systems. The differential equations valid for simple systems or in a laminar flow condition are shown to replicate self similarly on the emergent multifractal structures signaling onset of turbulence.

To explain  emergence of  anomalous multifractal scalings specifically 
in a turbulent plasma, we first discuss how the scaling of
a dynamical variable may become anomalous due to novel multifractal
contributions from the asymptotic boundary layer regions 
realized in a self consistent manner in a given fluid model. Such
anomalous scalings can be interpreted as the production in abundance of
spatio-temporal multifractal measures \cite{mandel} triggering turbulence
generating instabilities even in a simple electrostatic plane wave solution
of the kinetic Vlasov-Maxwell system \cite{plsm1}. As an another application
of this novel nonclassical measure, we next derive anomalous stretched
Gaussian scalings \cite{klafter,chen, plsm2} for the transport of charged
tracer particle distribution in a turbulent Vlasov plasma.

\section{Nonclassical Measures in Fluid}

Here, we present in brief the   mathematical arguments \cite{dp,dp1,dp3} leading to the emergence of
multifractal scalings  from  standard differential measures in a laminar flow 
as the original laminar flow tends to become turbulent. 
Recall that the traditional (differential) Lebesgue measure is well suited for simple systems, for
instance, the uniform rolling of a billiard ball along a straight line, say, or in a laminar flow.
In such a context if one envisages a limiting statement involving $%
x\rightarrow x_0$ ($x$ may denote the position of the ball), then it is reasonably 
correct to treat  this motion using the ordinary measure $\Delta x=x-x_0$
which goes to zero linearly with uniform rate 1. Motion
of a single fluid particle in a moving or static body of fluid medium
(Lagrangian view), however, is likely to present a new caveat. As the fluid
particle moves uniformly along a streamline (actually a path line) of a
laminar flow of the fluid medium, the standard \emph{Lebesgue measure}
apparently has to be modified because of a \emph{possible back reaction}
generated by the intervening fluid particles in the medium. In a laminar flow 
such back reactions would of course be negligible and can safely be ignored. 
However, in a turbulent flow the said back reactions from intervening fluid particles 
(molecules) need not be negligible. The line segment
joining any two points along a turbulent  path line, $x$ and $x_0$ say, would develop
more and more into an irregular curve as the particle at position $x$
is transported {\em non-uniformly} not only linearly but also through infinitesimal 
 \emph{smooth } jumps avoiding singularities those are created dynamically by the 
enhanced pressure generated within the flow. The length of the dynamically generated irregular curve in the intervening medium, which would grow indefinitely, would, however,  contribute a
finite (renormalized) correction term to the ordinary Lebesgue measure akin
to the renormalization group \cite{golden} actions delivering a \emph{finite observable
effect} from an \emph{apparent divergence} (Fig.1). More importantly, \emph{this
accumulation of an extra measure is realized in a manner consistent with the
original dynamical constraints of the motion}. 
%\begin{center}
\begin{figure}[!tbp]
%\caption{h} 
\caption{\small \sl \sl Nonlinear Limit: As a variable $t$ in fluid model approaches 0, the length of invisible (inaccessible)) sector $(0,t)$ diverges intermittently as $t\times t^{-\nu(t)}$. In the limit $t\rightarrow 0$, the neighbourhood of 0 in the visible (accessible) sector $t>0$ is  extended by a duality principle to a totally disconnected multifractal set and the original variable $t$ is replaced by the effective renormalized variable $t^{\alpha}$, where the multifractal scaling exponent $\alpha=\alpha(\eta)$ corresponds to the singularity spectrum of the associated multifractal measure. The effective variable $t^{\alpha}$ does not vanish even when $t$ goes to zero (see text for details).}
\centering
\includegraphics[scale=.5]{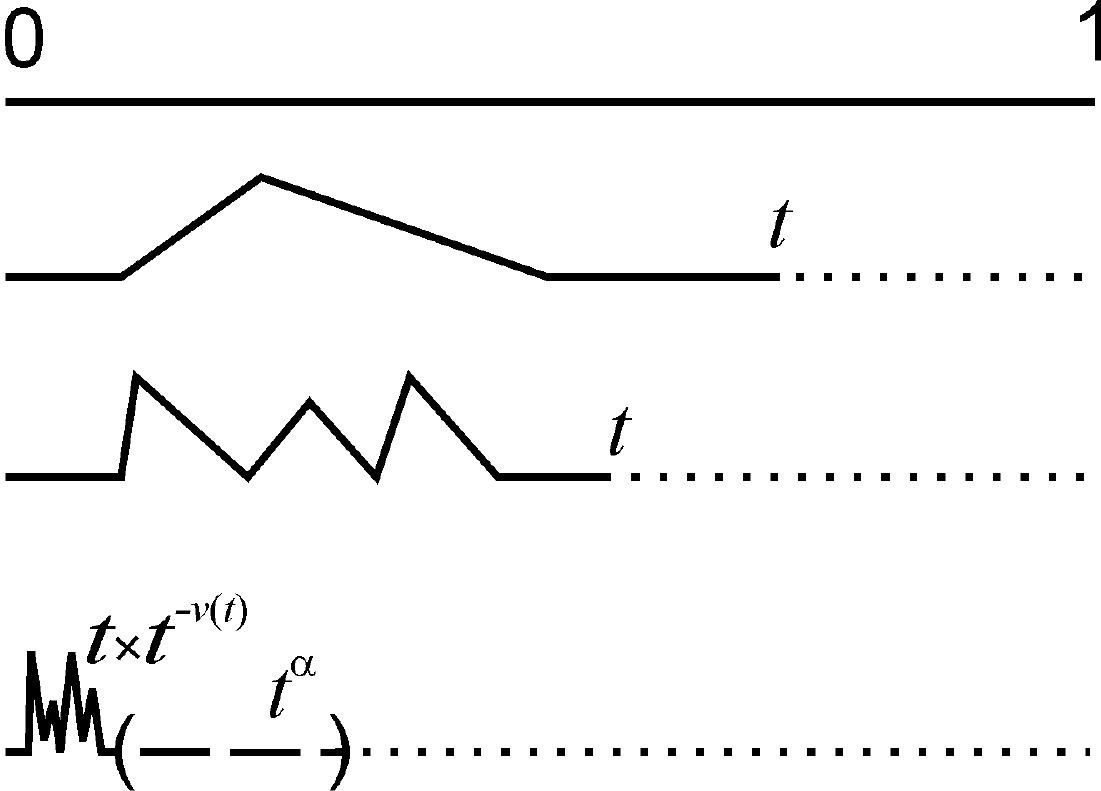}
\end{figure}
%\end{center}

\subsection{First Derivation}

To explain the above mentioned emergence of the nonclassical measure, let us
suppose that the dynamic variable $t$, greater than 1 initially, approaches 0 from the right. For definiteness we assume that the variable $t$ is dimensionless in the sense that $t$ satisfies the scale invariant equation 
\begin{equation}\label{sfe}
t\frac{dx}{dt}=x,\ x(1)=1  
\end{equation}%
The variable $x$ here may denote the position of a test particle in a uniformly flowing fluid.
Assume also that the real number line has a \emph{soft, fluid like structure} (one dimensional and may be
static) so that as $t>0$ approaches 0 the segment between 0 and $t$ gets 
\emph{squeezed} in the form of an irregular curve. The length of this
segment would grow indefinitely, supposedly in the form $t\times t^{-\nu(t)}
$, as $t$ closes gradually toward 0 and continues to become vanishingly
small, but does not vanish exactly (0 being a singular point, that may be
present in a specific physical problem or fabricated dynamically, see below). As long as
$t$ approaches 0 linearly as in an uniform laminar flow the factor $t^{-\nu(t)}$
would remain constant ($\approx 1$) so that $\nu(t)$ 
satisfies the scale invariant equation 
\begin{equation}\label{sfe2}
\log t\frac{d\nu}{d\log t}=-\nu
\end{equation}
and $\nu(t)=\frac{k_0}{\log t}$, $k_0>\approx 0$ being a constant.
However, as the line segment $(0,t)$ gets gradually squeezed due to effective 
compressive force generated by the nonlinear flow, we expect formation of a countable set of singular points 
where $\nu(t)$ is nondifferentiable in the usual sense. The length (measure) of the dynamically 
generated irregular curve inside  the interval $(0,t)$ would, however, grow gradually with the production of
 more and more  singularities in the interior. In the present approach 
such a possibility of dynamic production of a singularity set is  {\em avoided}, so to speak, by allowing the flow variable 
$t$  to execute smooth smaller scale jumps respecting eq(\ref{sfe2}) and its dual eq({\ref{sfe3}}). 

First, let us establish the successive growth of length of the interval $(0,t)$ with the generation of $n$ number of singular points $t_i: \ i=1,2,\ldots$ such that $0<t_{i+1}<t_i<t$. To begin with, suppose $t_1$ be the first singularity produced as the line segment $(0,t)$ is squeezed slightly. Exploiting the scale invariance, the exponent $\nu(t)$ would now satisfy eq(\ref{sfe2}) with $t$ replaced by the rescaled variable $\tilde t=t/t_1>1$ so that we  have $\nu(t)=\frac{k_1}{\log \tilde t}$, $k_1>k_0$, and the length of the intervening fluid line would grow as $te^{k_1}>t$. At the next level of squeezing, one considers generation of two singular points $0<t_2<t_1<t$. Proceeding exactly as above with the rescaled variable $t\mapsto \tilde{\tilde t}=t/t_2$ instead, one now arrives at an enhanced length  of the form $t\times t^{-\nu}=te^{k_2}, \ k_2>k_1$. Notice that $k_2<k_1$ is inconsistent with the proposed scenario. Thus with the production of more and more number of singular points the linear measure $t$ of the original segment $(0,t)$ would continue to grow multiplicatively as $te^{k_n}$, $k_n\rightarrow \infty$ as $n\rightarrow \infty$, when the $n$th level singularity $t_n$ is mapped  to $t=1$ by rescaling $t/t_n$. It follows, consequently, that the nature of the {\em generic} singularity at $t=1$  should alter successively as singularities from far deeper regions of $(0,t)$ are mapped to $t=1$ by rescalings. To conclude, the diverging growth in measure in the interior of the fluid segment $(-t,t)$ could be modeled by the simple, logarithmically scale invariant eq(\ref{sfe2}). The scaling exponent $\nu(t)=\frac{k_n}{\log {t^{\prime}}}>1, \ t^{\prime}=t/t_n>1$ would diverge in an intermittent manner as new singularity at $t_{n+1}$ is met as $t\rightarrow 0$ by squeezing. This dynamical emergence of singularity set and the associated diverging growth in measure is, however, eliminated (or avoided) by invoking {\em the principle of duality and inversion mediated jump modes} as outlined below.

To see explicitly this avoidance of singularity, suppose $t=1$ is the dynamically generated 
singularity as $t\rightarrow 0$ nonuniformly as described above. As noted already, singularity at any arbitrary point $0<t_0<t$ 
can be mapped to $t=1$ by scale invariance, so that the description of jump mode in the neighbourhood 
of $t=1$ can be taken as generic. In the said neighbourhood of the singular point 1, let us consider two 
unevenly distributed points $t_- <1< t_+$, $t_+=1+\eta$ and $t_-=1-g(\eta), \ g(\eta)\gtrapprox 0$ for $\eta\gtrapprox 0$ with the  property that $0<g(\eta)<\eta$. By transition via {\em inversion induced small scale jump} we mean $t_+\mapsto t_+^{-1}=t_-$, so that the exponent $\nu(t_+)\mapsto \nu(t_-)=\nu(t_+^{-1})\propto 1/\nu(t_+)$. As a consequence eq(\ref{sfe2}) valid in the right neighbourhood of 1 gets transformed into a {\em dually (inversely) related self similar replica}
\begin{equation}\label{sfe3}
\log t\frac{d\nu}{d\log t}=-\nu \  \longmapsto \ \log t\frac{d\nu}{d\log t}=\nu
\end{equation}
leading to $\nu(t_-)= k\log t_-^{-1}, \ k>0$. Notice that had the point $t=1$ been regular, eq(\ref{sfe2}) would have been extended over the left neighbourhood as well. To emphasize, we see that the {\em dual} eq(\ref{sfe3}) is essentially {\em the self similar replication of the original eq(\ref{sfe2}) by inversion in classically forbidden (that is to say, inaccessible) region of the turbulent flow.} 

In presence of this nontrivial jump mode, we now see that the effective scaling factor $t_+^{-\nu(t_+)}$ that remains {\em locally constant} in the right hand side of the singular point $t=1$ would make a smooth transition to a fluctuating variable $\tau=t_-^{-\nu(t_-)}$ by inversion induced jump. As $t_+\rightarrow 1$, the locally constant scale factor becomes a fluctuating variable $\tau\approx t_-^{-kg(\eta)}=t^{-\frac{k(g(\eta))^2}{\log t^{-1}}}, \ t\approx 0$ for a class of scaling functions $g(\eta)=\eta^s, \  \eta>0$. The exponent $0<s<1$ depends on the nature of the singularity point. Since there could at least be  a countable set of singularities in the interval $(0,t)$ for any small but finite value of $t$, $g(\eta)$ actually denotes a multifractal scaling function. As a consequence, in the limit $t\rightarrow 0$, the ordinary laminar flow measure $dt$ is transformed into the smooth multifractal measure $dt^{\alpha(\eta)}$ where $0<\alpha(\eta)=1-\frac{k(g(\eta))^2}{\log t^{-1}}<1$ varies over the singularity set of $(0,t)$. As, in the limit $t\rightarrow 0$, there are an uncountable number of possible singularity set (essentially, a family of  Cantor sets), exponent $\alpha$ represents a huge spectrum of turbulent fluctuations. It should be clear that the exponent $\alpha(\eta)$ has intermittent behaviour being almost constant slightly away i.e. as $g(\eta)\rightarrow \eta$ (so that $s\rightarrow 1$ and $\alpha(\eta)\approx 1$ as $\frac{k\eta^2}{\log t^{-1}}\approx 0$ relative to the first order infinitesimal $\eta$), from the singular point.

We remark that the dimensionless variable $t$ may either be the time variable or a space variable. The turbulent differential measure for a space variable would scale as $dx^{\beta(\eta)}$ for a multifractal exponent $0<\beta<1$.  

{\em Relationship with classical multifractal measure:}   A multifractal set is a set that is composed with a multitude of interwoven fractal subsets, each with differing fractal dimension \cite{fris}. The fractal dimension $D$ gives the idea of scaling of a measure $\mu$ with support on the fractal set $F$ with the size $\delta$ of a ball $B_x(\delta)$ centered at $x\in F\cap B_x(\delta)$: $\mu(B_x(\delta))\sim \delta^{-D}$. The fractal dimension $D$ is defined globally as a constant for a fractal set. For a multifractal the above scaling can be valid only locally:  $\mu(B_x(\delta))\sim \delta^{\epsilon(x)}$. Here, the exponent $\epsilon(x)$ denotes the {\em singularity strength} at the point $x$ and does not, in general, define the fractal dimension of any set. To generalize the concept of dimension for a multifractal set, there exist two approaches, related, in a sense, by a Legendre transform. The first approach starts by covering the support of multifractal measure by boxes of size $\delta$. Let $N_{\epsilon}(\delta)$ be the number of boxes that scales like $\delta^\epsilon$. Then, under mild restrictions \cite{falc}, it can be shown that there exists a convex function $\alpha(\epsilon)$ such that $N_{\epsilon}(\delta)\sim \delta^{-\alpha(\epsilon)} $  and the exponent  $\alpha(\epsilon)$, called the {\em singularity spectrum} of the multifractal measure, has the interpretation of the fractal (box or Hausdorff) dimension of an interwoven set of points $x$ having the singularity strength $\epsilon(x)=\epsilon$. 

In an alternative approach, one considers the partition function $S_{\delta}(q)=\sum_i \mu(B_i(\delta))^q$ for the support of the probability measure $\mu$ that is covered by the balls $B_i$. Again, under the stated mild conditions, the partition function has a power law scaling $S_{\delta}(q)\sim \delta^{\tau(q)}$ for $\delta\rightarrow 0$ and one arrives at  the generalized Renyi dimensions  $D_q=\tau(q)/(q-1)$. Further, the {\em generalized dimension spectrum} $\tau(q)$ relates to the singularity spectrum $\alpha(\epsilon)$ by the Legendre transform $\tau(q)=\underset{\epsilon}{\inf}{(q\epsilon-\alpha(\epsilon))}$. 

To interpret the smooth multifractal scaling exponent $\alpha(\eta)$, derived above from the inversion mediated jump modes, as the singularity spectrum of the dynamically generated multifractal measure in the soft model we proceed as follows. The neighbourhood of the generic singularity at $t=1$ is extended, by rescalings, into {\em non-classical} multifractal set that is covered by open intervals (balls) of size $\delta$. The measure concentrated on such a ball scales locally as $\delta^{\eta(\tau)}$. Here, the rescaled variable $\tau$ belongs to the multifractal set and relates logarithmically to the original variable $t$ in the vicinity of the generic singularity.  The singularity parameter $\eta(\tau)$ gives an estimate of the relative size of miniscule connected segments over a distribution of  miniscule open gaps in the neighbourhood of the point $\tau$ (c.f. Sec.2.3). The exponent $\alpha(\eta)$ now have the meaning of the {\em singularity spectrum} of the underlying multifractal measure that is generated dynamically from the diverging linear measure in the invisible (inaccessible) sector in the neighbourhood of the singularity at $t=1$  by the duality principle enunciated above.  We shall, however, continue to call the exponents $\alpha(\eta)$ and $\beta(\eta)$ loosely as multifractal scaling exponents in what follows.

\subsection{Alternative derivation}
 We give here another derivation of the  jump mediated multifractal increments to explain the origin of turbulent like behaviour in an asymptotic time as $t\rightarrow \infty$. This also points out the appearance of smaller and larger scales dynamically  thus making room for activation of jump modes in an otherwise featureless flow. Since we are concerned with
realizing a (scale invariant) nonlinear increment in a fluid medium from a linear laminar flow, 
 let us consider the simplest \emph{scale invariant} uniformly accelerated flow equation 
\begin{equation}
t\frac{dv}{dt}=v,\ v(1)=1  \label{sfe4}
\end{equation}%
in a bounded region of the form $\epsilon <t<\epsilon ^{-1}$ where $\epsilon 
$ may relate to a sufficiently distant time scale $T=\epsilon ^{-1/a},\
0<a=a(\epsilon )<1$. The exponent $a$ is reminiscent of an anomalous scaling
that will be justified below. In a simple, laminar like flow condition, the
conventional linear velocity increment of the form $\Delta v=v(t+h)-v(t)$ works
perfectly well in a fluid model and we also have $a=1$. However, a turbulent
like condition may be mimicked here even by eq(\ref{sfe4}) when one considers a
{\em singular limiting problem} defined by the concomitant limit $t\rightarrow
\epsilon ^{-1}$, but $\epsilon \rightarrow 0$ satisfying the condition $%
0<\epsilon <t^{-1}$. This singular problem is indeed \emph{nonclassical} because
of availability of \emph{infinitely large} scales of the form $\tau >1$,
when the classical analysis can have only to ordinary scales $1\leq
t<1/\epsilon $. In the present nonclassical approach, the point $\tau
=\epsilon t=1$ is a singularity that is unattainable classically as $%
t\rightarrow \infty $. However, in the extended scenario, increments in a
deleted neighbourhood of $\tau =1$ can be realized by an inversion of the
form 
\begin{equation}\label{inv}
\tau _{-}=\epsilon t\ (<1)\mapsto \tau _{-}^{-1}=\tau _{+}=t^{-\alpha (\epsilon)}  
\end{equation}%
where we set $\epsilon t^{a}=1$, in the limit $t\rightarrow \infty $ and $%
0<\alpha (\epsilon )=1-a(\epsilon )<1$, thus transferring an element at $%
\tau _{-}<1$ to $\tau _{+}>1$, on the extended rescaled real line,
instantaneously by the \emph{nonlinear but smooth} jump mode. More
importantly, above inversion induced scaling scenario provides an automatic
scheme for identifying the disconnected right neighbourhood of $\tau =1$ with
the right neighbourhood of $\tau =0$ of the real line $0<t<\epsilon ^{-1}$.
We rewrite this new rescaled inverted variable in the right neighbourhood of
0 by $t_{1}=t^{-\alpha (\epsilon )}$. Next, we notice that \emph{inversion
induced increments} as above satisfies eq(\ref{sfe4}), viz., $t\frac{dv}{dt}=v$, in the transformed
small scale monotonic variable $t_{1}=t^{-\alpha (\epsilon )}\gtrapprox 0$
when the ordinary differential increments are translation invariant. As a
consequence the inversion induced increments may be said to be translation
invariant in double logarithmic variable $\eta =\log \log t_{1}$ when
increments are taken as $\log h,\ 0<h<1$.

In a turbulent flow small/large scale nonlinear jump increments given by eq(%
\ref{inv}) are likely to play vital, predominant role over the ordinary
linear increments. Multiscale coherent structures such as eddies and
vortices generally endow a turbulent medium a multifractal structure leading
to nonuniform scaling exponents and intermittency. In the present approach
the jump mode scaling exponent $\alpha =\alpha (\epsilon )$ is \emph{not
necessarily a fixed constant} but a variable depending on the level of
singularity, that is to say, the regions or levels of scale changes as
encoded explicitly by $\epsilon $ and hence corresponds actually to a \emph{%
multifractal scaling} in the turbulent flow (c.f. Sec.2.1). More importantly, a fixed
exponent $\alpha $ can only be \emph{locally constant} in the sense that $%
\alpha $ is constant for the time scale $t_{1}<<\epsilon ^{-1}$ only, but
may vary in the next generation large scale $t_{1}\sim \epsilon ^{-1}$, and
so on. To see this explicitly, suppose $K$ is a constant i.e. $\frac{dK}{dt}%
=0$ for $\epsilon <t<<\epsilon ^{-1}$. However, as $t\rightarrow
T^{a}=\epsilon ^{-1}$ one gets $\epsilon \frac{dK}{d\epsilon t}=0$ and so by
letting $\epsilon \rightarrow 0$ one may deduce that $\frac{dK}{dt_{1}}\neq 0
$, when we make use of the jump mode transformations eq(\ref{inv}). So we
conclude that 
\begin{equation}
K=\mathrm{const.,\ for}\ \epsilon <t<\epsilon ^{-1};  \label{lc} \\
\mathrm{but}\ \frac{dK}{dt}\neq 0,\ \mathrm{when}\ t\gtrapprox 1/\epsilon 
\end{equation}%
because of the dynamical production of multifractal measures abundantly in
the asymptotic boundary layer regions.

Let us recall that in the literature of singular perturbation methods \cite{golden}, one considers a limiting 
ansatz of the form $(\tau_-=)\ \epsilon t=\epsilon^{-b}, \ b=b(\epsilon,t)$ which can be rewritten as 
$\tau_-=t^{\frac{b\log \epsilon^{-1}}{\log t}}$ so that under the proposed inversion mediated jump, one obtains 
$\tau_+=\tau_-^{-1}=t^{-\frac{\log t}{b\log \epsilon^{-1}}}$, which exactly matches  the derivations of Sec.2.1 
following the duality principle viz, eq(\ref{sfe3}), and also with eq(\ref{inv}) for
$\alpha(\epsilon)=\frac{\log t}{b\log \epsilon^{-1}}$. This also tells that the duality is already incorporated into the definition of 
$\alpha(\epsilon)$ (see below), so that explicit realization of the duality via eq(\ref{sfe3}) is not necessary in writing eq(\ref{inv}). In other words, the exponent $\alpha(\epsilon)$ is an effective renormalized quantity, invariant under the duality transformation.

Finally, to {\em justify} the {\em anomalous scaling} for $t$ as $t\rightarrow \infty $
in the present formalism let us remark that as $\eta =t^{-1}\rightarrow 0$
respecting $0<\tilde{\eta}_{n}<\epsilon ^{n}<t^{-1}$, relatively \emph{%
invisible} smaller scales $\tilde{\eta}_{n}$ residing in $(0,\epsilon ^{n})$
might have a coherent, cooperative\emph{\ effect} on the visible variable $\eta $ in
the form $\eta ^{-\alpha (\epsilon )}$ where the slowly varying, locally
constant effective exponent $\alpha (\epsilon )=\underset{n\rightarrow \infty }{\lim }%
\log _{\epsilon ^{-n}}(\epsilon ^{n}/\tilde{\eta}_{n})>0$ is interpreted as
an ultrametric valuation living in a multifractal set of microscopically
small and macroscopically large scales \cite{ds,dp1,dp3}. Clearly, cascades of infinitesimally small scale 
invisible elements $\tilde \eta_n$ are related {\em dually (i.e. inversely)} to a visible, limiting element $\eta$, 
and hence this definition of the scaling exponent $\alpha(\epsilon)$ naturally incorporates the duality transformation eq(\ref{sfe3}).  A complex turbulent flow is likely
to generate such a multifractal set abundantly. Notice that the valuation $\alpha(\epsilon)$ is nontrivial, 
even when the ordinary variable $\eta$ vanishes asymptotically and signifies cooperative effects of smaller
(invisible) scales in formation of complex structures by the inversion induced duality principle.
In Sec.2.1, dynamically generated  scaling exponent 
$\alpha(\epsilon)$ is interpreted as the singularity spectrum of the associated multifractal set.

To give an explicit example of the above remarks, let us suppose that the ordinary linear limiting variable 
$\eta\rightarrow 0$ acquires a nonlinear structure, of the form $\eta\mapsto \eta^\prime=\eta^{1-\alpha}$, by the limiting duality 
$\tilde\eta_n/\eta^{n}\propto (\eta/\eta^\prime)^n, \ n\rightarrow \infty$, so that $0<\alpha(\eta)=\frac{\log \eta/\tilde \eta}{\log \eta^{-1}}=\underset{n\rightarrow \infty}{\lim}\frac{\log \eta^n/\tilde \eta_n}{\log \eta^{-n}}<1$ where $\tilde \eta=\underset{n\rightarrow \infty}{\lim}\tilde \eta_n^{1/n}$, for the smaller scale cascade $0<\tilde\eta_n<\eta^n$. Nontriviality of the valuation  $\alpha(\eta)$ is verified by setting $\alpha(\eta)=\alpha_0$, a constant for all $\eta\in (0,1)$, so that $\alpha(\eta)$ trivially is an ultrametric. More generally, $\alpha(\eta)=\alpha_0\eta^{s(\eta)}$ is locally constant over countable open subintervals of (0,1) (i.e. $\eta^{s(\eta)}={\rm constant}$), with nontrivial variability at the points of (classical) discontinuity, as represented by the variable factor $\eta^{s(\eta)}$ with $s(\eta)=s$ remaining constant on an infinitesimally small subinterval of the set of variability of $\alpha(\eta)$ \cite{dp1,dp3}, reflecting the cooperative effects of smaller scale cascades.  The effective renormalized exponent $\alpha(\eta)$ thus has an intermittent behaviour as claimed. For a special set of values of the form $\eta=1/3^{m\log 3} , \  \tilde\eta=\eta\times 1/3^{m\log 2}, \ m>0$ , the scaling exponent $\alpha(\eta)=\frac{\log 2}{\log 3}$ represents dynamical formation of a monofractal, i.e. the classical triadic Cantor set of dimension $\alpha_0=\frac{\log 2}{\log 3}$ in the neighbourhood of the singular point 0. For a realistic nonlinear flow such a well patterned set of values are exceptional, leading instead to an intermittent  multifractal exponent $\alpha$ depending on  $\eta$. 

An important feature to the above derivations, as remarked already 
in Sec.2.1 is that the above multifractal
measure is valid both in the time and space domains. Ordinary Lebesgue
measure $dx$ (say) is now extended to the nonlinear multifractal measure of
the form $d_mx:=dX^{-1}=dx^{\beta(\epsilon )}, \ 0<\beta<1$, since in the far asymptotic limit considered
here linear increment $x \mapsto x+h$ gets extended to the nonlinear
increment $x\mapsto x+h^{\beta}$. Here, $X^{-1}=x^{\beta(\epsilon )}$ denotes the nonclassical
space variable extension (by inversion induced duality principle) in an arbitrarily small neighbourhood of $x=0$,
analogous to the time variable $T^{-1}$ near $t=0$. The neighbourhood of $x=0$ is said to have extended into a totally 
disconnected multifractal set since the exponent $\beta(\epsilon )$ is defined only locally and satisfies the inequality $0<\beta(\epsilon )<1$. An equation of the form eq(\ref{sfe}) on a connected line segment close to 0 would then be
replicated self similarly, by the said duality, on infinitesimally small disconnected line
segments in an appropriate nonclassical variable of the form $T^{-1}$ or in $%
X^{-1}$. The novel possibility of such nonlinear transitions would have
significance in offering new inputs in the study of instabilities and
turbulence.

Before closing this Section, let us point out that the production of
multifractal measures in abundance in both space and time coordinate spaces
in a fluid medium has an interesting physical (dynamical) interpretation. In
a high Reynolds number flow, the relative velocity between two contiguous fluid
elements is expected to be very high. As the separation (usual Lebesgue
sense) of the fluid elements is reduced to an infinitesimal level denoted by
the fiducial scale $\epsilon $, the diverging nonlinear measure, activated
in the intermediate fluid medium because of generation of very high
contractive forces by dragging effects in the high velocity flow, leave an
influence in the form of a finite effective multifractal measure in the
observable sector by inversion. The continuum fluid medium at the invisible
infinitesimal scales $<\ \epsilon $ appears to have assumed asymptotically
the granular structure with a diverging measure. The freedom of inversion
mediated transfer of enhanced measure, that is to say, {\em kinetic energy} from
the invisible sector to the observable dynamical sector $>\ \epsilon $ in
the present scenario creates the room for realizing new levels of
instabilities and complex flow patterns in the fluid. The present inversion
mediated energy enhancement is analogous in spirit of the renormalization
group techniques in critical phenomena and nonlinear systems \cite{golden}. The detailed
comparison of the two approaches in the fluid flow would be considered
separately.

\subsection{Activation of Jump Mode: Definitions}

Here we introduce the formal definition of the inversion induced jump modes. As before we consider two points $t_{\pm}=1\pm\eta, \ {\rm and} \ t_{+\mu}=1+\eta\mu(\eta), \ \eta<<1$ in the vicinity of $t=1$ which may or may not be a singularity for a flow. Further, let us we assume here that $t_-\mapsto t_-^{-1}=\lambda t_{+\mu}$, which is the definition of an {\em inversion induced jump} transferring the point at $t_-$ to a spectrum of possible points denoted by $t_{+\mu}$ and $\lambda$.  Our aim is to spell out the role of each of the parameters involved in the definition. We first note that $t_-^{-1}=1+\frac{\eta}{1-\eta}:=\tilde t_+ > t_+$, a unique point in the right hand side of $t=1$.  In case $t=1$ is regular,   all the above points attain the point $t=1$ as $O(\eta)$ when  $\eta\rightarrow 0$. The linear increment corresponding to inversion induced transition is  $\Delta_i t_-=t_-^{-1}-t_- =\frac{\eta(2-\eta)}{1-\eta}\sim O(\eta)$. Noting that {\em linear} increment for shift mediated flow from $t_-$ to $t_+$ is $\Delta t_-=t_+-t_-=2\eta$, the relative gain in the jump mediated increment over  the ordinary shift  defined by $\Delta_j t_-=(\Delta_i t_--\Delta t_-)/\Delta t_-=O(\eta)$  vanishes as $\eta\rightarrow 0$. As a consequence, in a regular flow the {\em measure of the activation of jump modes} over the linear mode, defined as the logarithm of the rate of variation of the relative gain of jump mediated increment over the usual linear increment, viz. $J=\log_{\Delta t_-}(\Delta_jt_-/\Delta t_-)$, vanishes in the limit $\eta\rightarrow 0$. Hence forth, we call $\Delta_j t$ the {\em pure jump increment} at the point $t$ and evaluated asymptotically in an open neighbourhood of the form $(t-t\eta,t+t\eta)$ when $\eta\rightarrow 0 \ (\neq 0)$.  

In the nontrivial case, when $t=1$ is a singularity, developed dynamically in a flow, the first order infinitesimal $\eta$ can approach 0 arbitrarily close but, nevertheless, fails to reach 0 exactly. In this case we demonstrate a nontrivial activation of a jump mode asymptotically as $\eta\rightarrow 0$. 
 
As $t_-\rightarrow 1$, nonlinear back reaction from  the singularity creates a room to invoke jump mode  
so that the dynamical variable $t$ at $\tilde t_-=1-\eta\mu(\eta) =t_-^{\mu(\eta)} +O(\eta^2), \ \mu(\eta)<1, \eta<<1$  takes a jump $\tilde t_-\mapsto \tilde t_-^{-1}\propto t_{+}$ to bypass the singularity and thus to allow the flow to continue smoothly but with added structures for $t>1$. As a consequence, the small scale variable $\eta$ in the right hand side of the singularity denotes a growing variable in the interval (0,1). The above relation can be rewritten in the equivalent form of a jump mode introduced previously (c.f. Sec.2.1):  $ t_-\mapsto t_-^{-1}= \lambda t_{+\mu^{-1}}+O(\eta^2)=\lambda (1+\eta\mu^{-1}(\eta))+O(\eta^2), \ \lambda<1$. The function $\mu$ is assumed to belong to the class of   sufficiently differentiable functions for $\eta>0$ that represents non-specificity or uncertainty in the exact  moment in realizing a jump before an emerging singularity. The proportionality parameter $\lambda$ denotes possible uncertainty in the final moment. The combined effect of uncertainty now is rewritten as $t_-^{-1}=t_+^{\nu} +O(\eta^2)$ where $\nu=\mu^{-1}+\frac{\log \lambda}{\log t_+}$. For definiteness we assume $\nu>1$. This corresponds to continuity of flow and associated structure formation induced by the singularity as referred to above. 

It now follows that the jump increment $\Delta_j t_-$, in the nontrivial case, has the form  $\Delta_j t_-= \frac{1}{2}(\nu-1)+O(\eta)$. In the presence of a singularity at $t=1$, one may imagine an open gap of the form $(1-\eta,1+\eta)$ that is crossed by a jump. The above {\em jump increment} then gives a {\em measure} of the  size of a compact connected segment  determined by $\nu$ relative to the open gap of size $2\eta$.  Because of intrinsic scale invariance of the jump mode (c.f. Sec. 2.1), a class of nontrivial jump modes is determined uniquely by $\nu=1+2\eta^{\alpha}, \ \eta>0$ where $0<\alpha(\eta)<1$. The associated nonlinear differential measure  for jump mediated flow is defined by $d_j t=d {\tilde t}^{\alpha(t)}$  where $\tilde t\approx 0$ is a scale invariant small scale growing variable in the neighbourhood of a singularity. As a consequence, the ordinary measure $dt$ gets extended to the nonclassical measure $dT$ for $T=t(1+{\tilde t}^{\alpha}/t)$ where the nontrivial part, representing smooth intermittent fluctuations, is activated only in a small neighbourhood of a singularity when $\tilde t\approx 0$ but never vanishes exactly.  It also follows that the corresponding {\em activation parameter} $J=\log_{\Delta t_-} \frac{1}{\Delta t_-}\Delta_j t_-=(1-\alpha), \ \alpha\neq 1$ relates to  the dimension of the multifractal  set on which  jump modes live.  This activation parameter has an equivalent ultrametric interpretation \cite{ds,dp1,dp3} and was referred to in Sec.2.2. 

To summarize, we have presented two different routes for derivation of multifractal differential measures from a soft (fluid) model of real line. It is argued that the linear Lebesgue  measure follows from the ordinary stiff model of real line and should apply or be relevant for laminar type flow only. In a complex turbulent flow Lebesgue measures of the form $dt$ and $dx$  must be extended to multifractal measures of the form $dt^{\alpha(\epsilon)}$ and $dx^{\beta(\epsilon)}$ signifying dynamical onset of multifractal sets in the turbulent flow. The exponents $\alpha(\epsilon)$ and  $\beta(\epsilon)$ correspond to the singularity spectra of the multifractal measures in the time and space domains respectively of the nonlinear flow. The first derivation of the multifractal measures makes use of a duality principle in the context of the scale invariant equations eq(\ref{sfe2}) and eq(\ref{sfe3}). The alternative approach is based on an effective valuation that is invariant under the jump mediated duality transformation. The formal definitions of jump increment and the associated jump differentials are presented in Sec.2.3. A measure called the activation parameter $J$ is also introduced, non-triviality of which demonstrates the onset of nonlinear incremental modes in the flow. It also follows that the jump increment for $t$ ( or $x$) in the neighbourhood of $t=1$ ( or $x=1$) corresponds to the ordinary increment for the scaling function $(\log t)^{\alpha(\epsilon)}$ ( or $(\log x)^{\beta(\epsilon)}$), $0<\alpha(\epsilon)<1 \ (0<\beta(\epsilon)<1))$.

\section{Applications to Plasma}

Here we present two applications of the above formalism in plasma
instabilities and plasma turbulence. These examples are expected to confirm
that the instabilities considered here actually belong to altogether a new
class of instabilities that plague any linear and nonlinear system. We
consider the well known plasma model described by the collision-less kinetic
Vlasov equation 
\begin{equation}
\frac{\partial f_{j}}{\partial t}+v_{i}\frac{\partial f_{j}}{\partial x_{i}}+%
\frac{q_{j}}{m_{j}}(\vec{E}+\vec{v}\times \vec{B})_{i}\frac{\partial f_{j}}{%
\partial v_{i}}=0  \label{vlas1}
\end{equation}%
where the distribution function $f_{j}$ describes the number of charged $%
q_{j}$ particles of species $j$ of mass $m_{j}$ with velocity $\vec{v}$ and
position $\vec{x}$ at time $t$. The electromagnetic fields $\vec{E}$ and $%
\vec{B}$ created by the plasma particles in the vicinity of $\vec{x}$ and at
time $t$ satisfy the full set of Maxwell's equations 
\begin{equation}
\nabla \cdot \vec{E}=4\pi \rho _{c},\ \ \ \nabla \times \vec{E}=-\frac{%
\partial \vec{B}}{\partial t},\ \ \ \nabla \cdot \vec{B}=0,\ \ \ \nabla
\times \vec{B}=\frac{\partial \vec{B}}{\partial t}+4\pi \vec{J}  \label{max}
\end{equation}%
where we use normalized electromagnetic units i.e. permittivity $\epsilon =1$%
, permeability $\mu =1$, velocity of light $c=1$. Further, charge and
current densities are defined self-consistently by the distribution
functions $f_{j}(\vec{x},\vec{v},t)$ as moments over velocity field $\rho
_{c}=\sum q_{j}\int f_{j}d^{3}v$ and $\vec{J}=\sum q_{j}\int \vec{v}%
f_{j}d^{3}v$. As a consequence, the Vlasov-Maxwell equations (\ref{vlas1})
and ({\ref{max}}) are truly nonlinear system of equations. The continuity
equation and the corresponding Navier-Stokes equation for the mean fluid
velocity field $\vec{u}=\int \vec{v}fd^{3}v_{i}$ viz, 
\begin{equation}
\frac{\partial \vec{u}}{\partial t}+\vec{u}\cdot \frac{\partial \vec{u}}{%
\partial \vec{x}}=-1/\rho \nabla \cdot \mathbf{P}-\rho _{c}(\vec{E}+\vec{u}%
\times \vec{B})  \label{ns}
\end{equation}%
follow easily from the kinetic Vlasov equation. Here, $\mathbf{P=\sum \int 
\vec{v}\otimes \vec{v}f_{j}d^{3}v}$ denotes the pressure tensor and $\rho $
is the mass density of the plasma fluid. It follows that the macroscopic
plasma flow is analogous to, but not exactly equivalent to the Navier-Stokes
flow of an ordinary fluid, because of the closure problem as reflected by
the occurrence of the pressure tensor.

\subsection{New Instability}

Let us consider the electrostatic wave described by the electrostatic
Vlasov-Poisson equation 
\begin{equation}
\frac{\partial f}{\partial t}+v_{i}\frac{\partial f}{\partial x_{i}}+\frac{q%
}{m_{e}}(\nabla _{i}\Phi )\frac{\partial f}{\partial v_{i}}=0  \label{vlas2}
\end{equation}%
where the electrostatic potential $\Phi $ satisfies the Poisson equation 
\begin{equation}
\nabla ^{2}\Phi =4\pi q[\int f(x_{i},v_{i},t)d^{3}v-\frac{\rho _{0}}{m_{i}}]
\label{poi}
\end{equation}%
Here we restrict $f$ to the one dimensional distribution function of
electrons and $\rho _{0}$ is the ion density considered constant. An
equilibrium state of this model is given by $f=f_{0}(v)$, and $E=0$ along
with the consistency condition $\int f_{0}(v)dv=\frac{\rho _{0}}{m_{i}}$. A
simple linear stability analysis for plane wave $\sim e^{i(-\omega t+kx)}$
for the first order fluctuations $f_{1}(x,v,t)$ and $\Phi _{1}(x,t)$ for
electron distribution and electric potential now leads in the long wave
length limit $k\rightarrow 0$ the warm plasma Langmuir wave dispersion
relation \cite{plsm1} 
\begin{equation}
\omega ^{2}=\omega _{p}^{2}+3<v^{2}>k^{2}  \label{disp1}
\end{equation}%
where $\omega _{p}^{2}=4\pi \rho _{0}q^{2}/m^{2}$ is the characteristic
plasma frequency and $<v^{2}>=\int v^{2}f_{0}(v)dv$ is the thermal spread
due to pressure.

Such linear wave stability may, however, experience \emph{new level of
instabilities} triggered by \emph{influx of energy transfer} by inversion
from smaller and larger scales available in the present formalism. We recall
that the Vlasov model is a collisionless limit of Boltzmann equation and is
valid for a plasma where collective effects dominate over the collisional
effects. Stated in time domain, this translates to the condition that the
relevant time scale $t$ must be $<$ the collision time scale $t_{c}$. As a
consequence the validity region of the Vlasov equation may be safely
considered to be $\epsilon <t<\epsilon ^{-1}$ where $t_{c}=\epsilon
^{-1/a(\epsilon )},\ 0<a(\epsilon )<1$. Because of translation invariance of
the Vlasov equation eq(\ref{vlas2}), the initial equilibrium state may be
realized at $t_{0}=\epsilon $. However, appearance of a fiducial parameter $%
\epsilon >0$ is significant in producing smaller and large scales having
nontrivial dynamical implications as noted above.

As explained in Sec. 2, the presence of the fiducial scale $\epsilon $ now
allows one to introduce rescalings $t\mapsto \tilde{t}=\epsilon t$ and $%
x\mapsto \tilde{x}=x/\epsilon $ which should leave the
Vlasov equation (\ref{vlas2}) invariant. As $t\rightarrow \epsilon ^{-1/a(\epsilon)}$
and $\Delta \tilde{x}=\Delta x/\epsilon ^{1/b(\epsilon )}\sim 1,\
0<b(\epsilon )<1$, one then obtains the following transformed differential
measure $d\tilde{t}=dt^{\alpha (\epsilon )}$ and $d\tilde{x}=dx^{\beta
(\epsilon )},\ \beta (\epsilon )=1-b(\epsilon )$ for the time and space
scaling exponents $0<\alpha (\epsilon )<1$ and $0<\beta (\epsilon )<1$. The
original Vlasov-Poisson's equation thus is replicated self similarly by
inversions on microscopic time and space scales, the infinite hierarchy of
such system of equations is written in a succinct manner as 
\begin{equation}\label{vlas3}
\frac{\partial f_{1}}{\partial \tilde{t}}+\tilde{v}_{i}\frac{\partial f_{1}}{%
\partial \tilde{x}_{i}}+\frac{\tilde{q}}{\tilde m_{e}}(\nabla _{i}\tilde \Phi _{1})\frac{%
\partial f_{0}}{\partial \tilde{v}_{i}}=0  
\end{equation}%
and 
\begin{equation} \label{poi2}
\nabla ^{2}\tilde\Phi _{1}=4\pi \tilde{q}\int f_{1}(\tilde x_{i},\tilde v_{i},\tilde t)d\tilde v  
\end{equation}%
using effective scaling variables $\tilde t=t^{\alpha(\epsilon)}$ and $\tilde x=x^{\beta(\epsilon)}$, w
here $\tilde{v}_{i}=\epsilon ^{-2}v,\ \tilde{\Phi}=\epsilon ^{-5}\Phi $, $\tilde{q}=\epsilon ^{-1}{q}$ and $\tilde{m}_{e}=\epsilon ^{-2}m_{e}$ are the
transformed velocity, field potential, electron charge and mass
respectively, leaving the plasma frequency $\omega _{p}$ scaling invariant.
The plane wave solution in the transformed variables of the form $\propto \
e^{i(-\tilde{\omega}\tilde{t}+\tilde{k}\tilde{x})}$ would then have the
dispersion relation for corresponding transformed frequency $\tilde{\omega}$
and wave number $\tilde{k}$: 
\begin{equation}
\tilde{\omega}^{2}=\omega _{p}^{2}+3<\tilde{v}^{2}>\tilde{k}^{2}
\label{disp1}
\end{equation}%
where $\tilde{\omega}=\omega ^{\alpha (\epsilon )}$ and $\tilde{k}=k^{\beta
(\epsilon )}$ which follow from dimensional analysis. As a consequence one
obtains finally $\omega ^{2}=(\omega _{p}^{2}+3<\tilde{v}^{2}>k^{2\beta
(\epsilon )})^{1/\alpha (\epsilon )}$. New sources of instabilities are
manifested from the fact that the exponents are noninteger $0<\alpha ,\
\beta <1$ and $-\infty <k<\infty $. For example, for exponents $\alpha =1/3$
and $\beta =1/2$ instability sets in for $k<-\omega _{p}^{2}/3<\tilde v^2>$. Moreover,
infinite set of stable and unstable modes are possible when either $\beta $
or both the exponents are irrational. To consider an example, let $\alpha$ be rational  
and $\beta$ irrational. Then infinite number of stable and unstable modes are given by 
\begin{equation}\label{irr}
\Re \ \omega_n^{2\alpha}=\omega_p^2+3<\tilde v^2>|k|^{2\beta}\cos 2\beta\pi n,  \ \
\Im\  \omega_n^{2\alpha}=3<\tilde v^2>|k|^{2\beta}\sin 2\beta\pi n
\end{equation}
For $\alpha=1/2$, in particular, we have stable but de-attenuated modes for $\sin 2\beta\pi n<0$. The modes are unstable and attenuated when the inequality is reversed. The plane wave profile 
\begin{equation}\label{wave}
e^{i(-|{\omega}^{\alpha}|{\omega}^{\alpha}T+ |k|^{\beta} k^{\beta}X)}
\end{equation}
 in the normalized stretched coordinates $T=\frac{t^{\alpha}}{|\omega|^{\alpha}}, \  X=\frac{x^{\beta}}{|k|^{\beta}}$ would explicitly display
 both these attenuations and instabilities activated by irrational scaling exponent in space coordinate (Fig.2 \& Fig.3). 
The generation of such a
large number of stable and unstable modes is triggered generically by the
dynamic creation of {\em localized} multifractal structures with irrational exponents in the
spatio-temporal configuration space of the plasma medium. The cooperative
effect of these irrational modes is expected to induce a complex turbulent behaviour
even in an otherwise stable electrostatic wave profile.

\begin{figure}[!tbp]
%\caption{h} 
\caption{\small \sl \sl  Plane Cosine wave $z=Cos (x-\sqrt 2 t)$ and Stable but de-attenuated irrational Cosine wave for $n=1$ in stretched variables with $\alpha=1/2, \ \beta=1/\sqrt 2$ and $k=1$, $\omega_p=1$, $3<\tilde v^2>=1$.} 

\includegraphics[scale=.5]{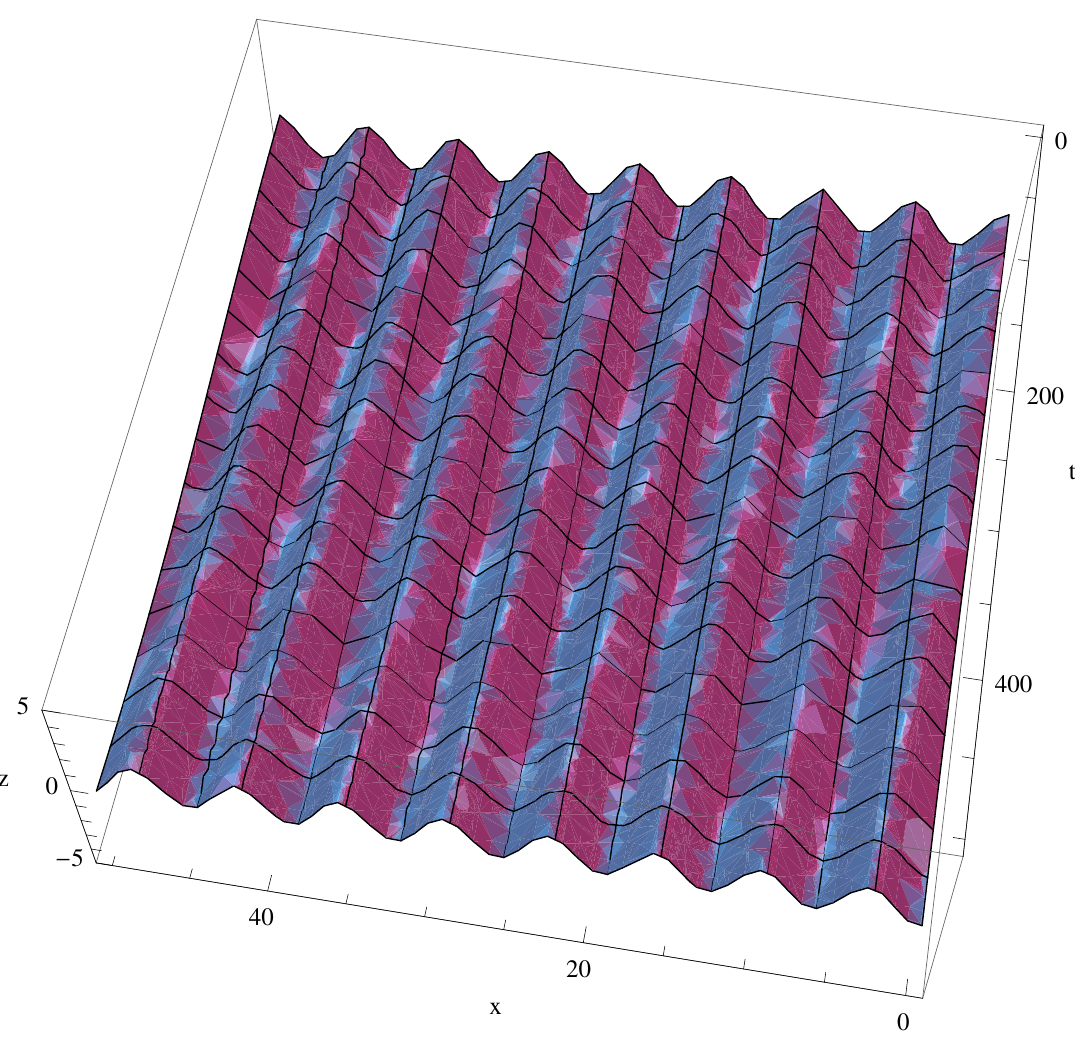}
\hspace{.5in}
\includegraphics[scale=.5]{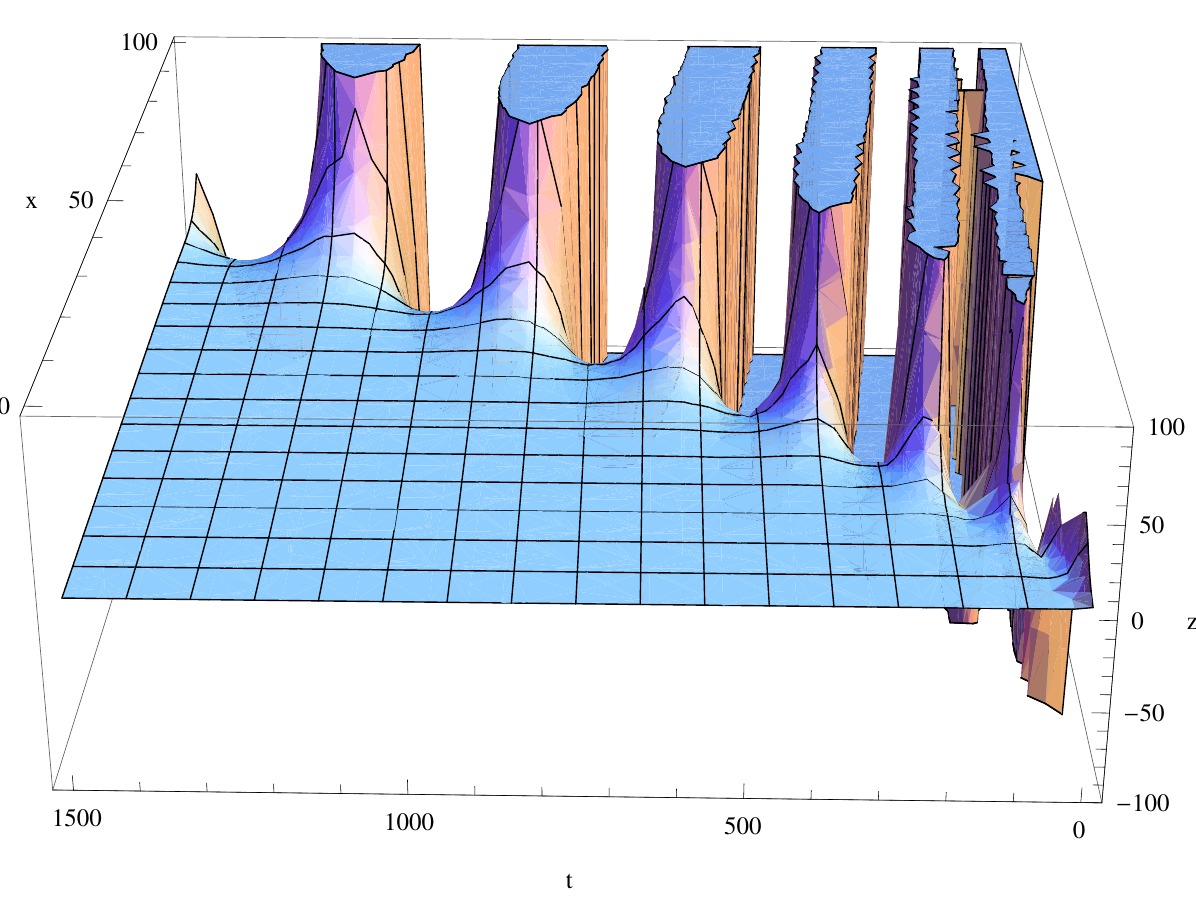}
\caption{\small \sl \sl  Unstable, attenuated irrational Cosine wave for $n=2$ in stretched variables with $\alpha=1/2, \ \beta=1/\sqrt 2$ and $k=1$, $\omega_p=1$, $3<\tilde v^2>=1$.}
\centering
\includegraphics[scale=.5]{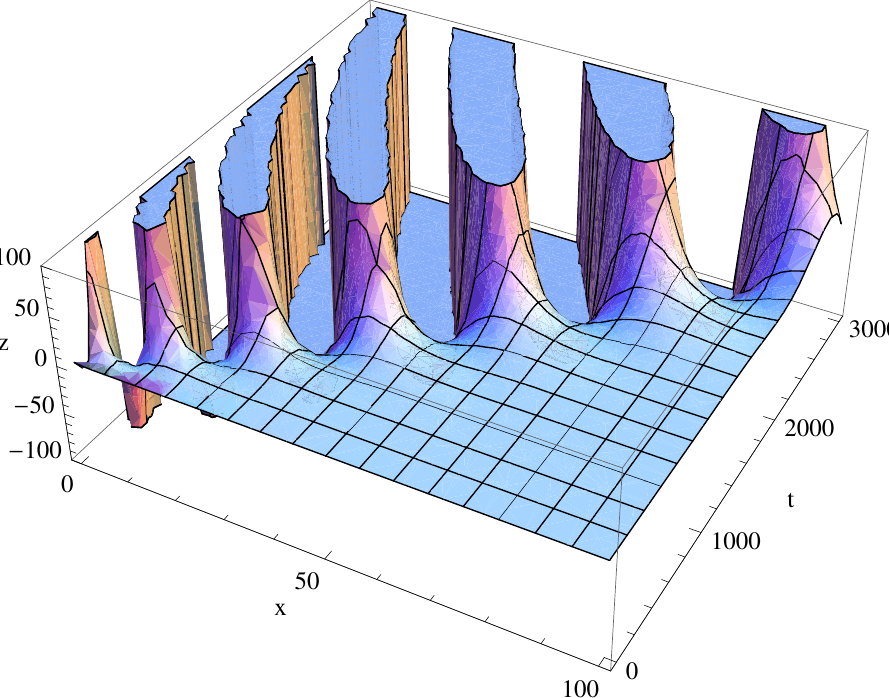}\label{a}  

\end{figure}

To conclude this section, we note that \emph{turbulence is expected to set
in much early in the present scenario compared to that from the conventional
approach in an electrostatic plane electron wave}.

\subsection{Anomalous Transport}

Our objective here is to study (charged) test (tracer) particles
distribution within a model of the Vlasov plasma in the present dynamical
scenario. As noted already, the Vlasov equation eq(\ref{vlas2}) is valid
till the collective mode time scale $t$ is much smaller than the collision
time scale $t_{c}$. In the above section we have seen that turbulence may
set in even in a time scale $t\sim t_{c}^{a}<<t_{c}$ for a small
multifractal exponent $0<a=a(\epsilon)<<1$. Recalling the fact that the collisionless
equation (\ref{vlas2}) is nothing but the conservation of number density
along a particle trajectory in the phase space i.e. $\frac{df}{dt}=0$, the
onset of turbulence can also be associated with the loss of conservation of
number density i.e. $\frac{df}{dt}\neq 0$ when $t\sim t_{c}^{a}$ mimicking
collisional contributions from multifractal measures generated in the
asymptotic boundary layer regions (c.f. eq(\ref{lc}) ).

Let us note next that with the onset of turbulence in, for example, the
electrostatic model, the induced magnetic field fluctuations $\vec{B}$ need
not be negligible, even in the simplified situation when one could continue
to neglect the spontaneously enhanced collisional effects. The electrostatic
model of Sec.3.1 now would be extended to the Vlasov-Maxwell equations (\ref%
{vlas1}) and (\ref{max}). The test particle trajectory under the $\vec{E}%
\times \vec{B}$ drift wave limit is then given by 
\begin{equation}
\frac{d\vec{x}}{dt}=\frac{1}{B^{2}}\vec{E}\times \vec{B}  \label{test}
\end{equation}%
Because of strong nonlinear coupling in the plasma model in presence of
turbulence generating instabilities, a statistical treatment of this
equation is considered appropriate. Any self-consistent solution of the
plasma model would realize the electromagnetic field vectors $\vec{E}$ and $%
\vec{B}$ as random fluctuating fields, so that the tracer test particle
equation (\ref{test}) is interpreted as a stochastic Langevin equation.

As a specific simplified model, suppose the tracer particle travels
initially in a steady electrostatic plasma fluid governed by the
Vlasov-Poisson equation and $\vec{B_{0}}$ is a constant magnetic field along
the $z$-axis, say. In a steady, stationary fluid profile, the tracer
trajectory is likely to resemble a Brownian path, so that the underlying
stochastic process driving the tracer particle equation (\ref{test}) is
Gaussian. The corresponding macroscopic description of this Brownian motion
is given by the advection-diffusion (AVD) equation 
\begin{equation}
\frac{\partial C}{\partial {t}}+{u}_{i}\frac{\partial C}{%
\partial {x}_{i}}={\frac{\partial }{\partial x_i}}{D\frac{\partial C}{\partial {x}_{i}}}
\label{advdiff1}
\end{equation}%
where $D$ is diffusivity and $C(x_{i},t)$ is the concentration density of an
ensemble of tracer particles. Under the Markovian approximation $D=\underset{%
t\rightarrow \infty }{\lim }\frac{\sigma ^{2}(t)}{2t}$ is approximately a
constant leading to Gaussian scaling for the mean square deviation $\sigma
^{2}(t)\sim 2Dt$.

Such a Gaussian scaling for particle trajectory however is expected for very
short time evolution of the model. For a sufficiently large time i.e. when $%
t\sim T>>1$, we rewrite (\ref{advdiff1}) by making use of its rescaling
symmetry via the rescaled variables $\tilde{t}=\epsilon t$, $\tilde{x}%
_{i}=x_{i}/\xi (\epsilon ),\ \epsilon ^{-1}=T^{a}(\epsilon )\ 0<a(\epsilon
)<<1$ 
\begin{equation}
\frac{\partial C}{\partial \tilde{t}}+\tilde{u}_{i}\frac{\partial C}{%
\partial \tilde{x}_{i}}=\nu \frac{\partial ^{2}C}{\partial \tilde{x}_{i}^{2}}
\label{advdiff2}
\end{equation}%
where the rescaled diffusivity $\nu =D/\epsilon ^{1+2h(\epsilon )}$ can
become quite large when $\xi ^{2}\sim \epsilon ^{2h(\epsilon )},\
0<h(\epsilon )<1/2$.

Next, we observe that the rescaled mean velocity of the fluid $\tilde u_i=%
\frac{d\tilde x^i}{d\tilde t}=\epsilon^{-(1-h)}u_i>>u_i$. The Reynolds number
corresponding to the late time evolution of the plasma fluid would therefore
be quite large, signaling once again the onset of turbulent inducing
instabilities in the plasma fluid. The equation (\ref{advdiff2}) would
therefore represent the turbulent advection-diffusion equation. \emph{%
Although looks apparently Gaussian, the underlying motion in the rescaled
variables is actually non-Gaussian, that we now establish}.

We proceed in steps following the scenario presented in Sec.2:

1. We first \emph{recall} that the rescaled variables actually denote \emph{%
nontrivial inversion induced transformations} (c.f. Sec.2) in the late
asymptotic limits $t\rightarrow \infty $ and $\epsilon \rightarrow 0^{+}$,
in the sense that the boundary point $\tilde{t}=1$ is singular and
unreachable; however the singularity is avoided by inversions of the form $%
\tilde{t}\mapsto {\tilde{t}}^{-1}=(\log \tilde{t})^{\alpha (\epsilon )},\
0<\alpha (\epsilon )<1$, where $\tilde{t}$, in the r.h.s, lies close to 1,
but exceeds 1.

2. Points 0 and $\infty $ of the positive real line are identified in a
nontrivial manner under the above inversion. The asymptotic limit generates
countable numbers of disconnected self similar sectors in the neighbourhoods
of $\tilde{t}=0$ and $\tilde{t}=1$, with the identification of the right
neighbourhood of $\tilde{t}=0$ of sector 1 with that of the right
neighbourhood of $\tilde{t}=1$ of sector 2, and so on. \emph{The system of
plasma (Vlasov-Maxwell, in the present case) equations are proliferated self
similarly over each of the neighbourhoods.} One particular example is
presented in Sec.3.1 for the electrostatic model. As a hindsight, \emph{the
neighbourhood of the origin of the physical coordinate space acquires a
granular structure which in turn is mapped nontrivially to infinity by
inversion}. Physically, this mean that new sources of kinetic energy become
available by inversion from the asymptotic boundary layers (c.f. final
remarks in Sec.2).

3. At a macroscopic level \emph{such generation and proliferation of
underlying plasma model over infinite number of disconnected (granular)
asymptotic sectors would actually trigger formation of locally quasi-stable
coherent structures (eddies etc.) leading to a new route to turbulence}.

4. The asymptotic coherent structures shall produce effective (renormalized)
multifractal measures, replacing the ordinary laminar state measures $dt$
and $dx_{i}$ respectively by the multifractal turbulent, but nevertheless,
smooth measure $dt^{\alpha }$ and $dx_{i}^{\beta }$ where the multifractal
exponents $\beta=\beta(\epsilon) >0$ and $\alpha=\alpha(\epsilon) >0$ respectively represent 
localized spatio-temporal
scalings of asymptotic coherent structures (c.f. Sec.2).

5. For high Reynolds number $Re$, i.e. when $\nu \approx 0$ (equivalently
very high mean velocity) the original laminar flow enters into a turbulent
flow with a concomitant transformation of the laminar AVD equation (\ref%
{advdiff1}) for passive tracers approximately into the turbulent AVD
equation (\ref{advdiff2}) involving asymptotic rescaled variables $\tilde{t}$
and $\tilde{x}_{i}$. In the asymptotic limits $t\rightarrow 1/\epsilon $, $%
\Delta x_{i}\rightarrow \xi (\epsilon )$ and $\epsilon \rightarrow 0$, the
time and space variables are transformed into the smooth \emph{asymptotic
multifractal} scaling laws by inversion, viz; $d\tilde{t}\mapsto
dT^{-1}=dt^{\alpha (\epsilon )},\ \alpha (\epsilon )>0$ and $d\tilde{x}%
_{i}\mapsto dX_{i}^{-1}=dx_{i}^{\beta (\epsilon )},\ \beta (\epsilon )>0$ as
explained in Sec.2 (we assume, for simplicity, spatial homogeneity $\beta
_{i}=\beta ,\ \forall i$).

We recall that for a low Reynolds number (i.e. low mean speed) laminar flow
increments and differential measures follow the standard Lebesgue measures
of classical analysis. However, for higher Reynolds number flows tend to
inhabit eddies of all possible length scales between the integral scale and
the Kolmogorov's molecular diffusion (dissipative) scale (i.e. between the
largest and smallest \emph{eddy} sizes). The turbulent AVD equation (\ref%
{advdiff2}) is valid in the intermediate time $t_I<<t\sim 1/\epsilon$ and
space $l_K<<x_i<<l_I$ scales where subscripts $K$ and $I$ denote appropriate
Kolmogorov and integral scales respectively. Assuming that \emph{formation
of eddies represent nonlinear structure formations in space and time
dynamically}, the nonlinear multifractal spatial increments as derived in
Sec.2 provide a general framework of endowing a smooth, nevertheless,
nonlinear measures on \emph{a multifractal fluid medium analogous to and
paralleling the renormalization group analysis delivering a few finite
measurable quantities in the form of scaling exponents even in the absence
of a detailed microscopic theory of the given dynamical problem} (i.e. the
variation of tracer concentration in turbulent plasma flow). Here, the
exponent $\beta$ denotes one of the scaling parameters derived from a
general argument based on an \emph{inversion induced structure formation
scenario} bypassing a detailed theory of eddy formation and a specific model
dependent derivation of nonlinear increments. One recently fashionable
approach available in literature is the framework of fractional calculus. As
remarked already, our approach is independent of fractional calculus
techniques and as shown below can capture anomalous scaling of higher
moments of tracer distribution quite naturally and easily. On the time
domain, on the other hand, for a time scale greater than the integral scale $%
t_I$, the presence of infinitely large scales close to and beyond the time
scale defined by the Reynolds number $t_R\sim Re\sim\nu^{-1}$, motion of a
tracer particle will be enhanced indefinitely which will then contribute a
small measurable renormalized quantity in the form of the nonlinear measure $%
dt^{\alpha}, \ \alpha>0$ by inversion. For small positive values of $\alpha$
this nonlinear measure would model decelerated and/or sub-diffusive processes
when $\alpha>1$ can lead to accelerated advection and/or super-diffusion.

Now, recognizing eq(\ref{advdiff2}) as a standard AVD equation in the
nonlinear deformed (stretched) variables $\tilde{t}=t^{\alpha(\epsilon) },t>0$ and $%
\tilde{x}=|x|^{\beta(\epsilon) }$, the concentration density with an initial delta
pulse $C(x,0)=\delta (x)$ in a turbulent flow should have the generic form
of the normal Gaussian distribution $C(x,t)=(\frac{1}{4\pi \kappa \tilde{t}}%
)^{\frac{n}{2}}e^{-\frac{\tilde{X}^{2}}{4\nu \tilde{t}}}$ in the stretched
variables $\tilde{t}$, $|\tilde{X}|$ where $X=x-Vt$. We remark that the
definition of the stretched space variable $\tilde{X}$ takes care of the
concomitant change in dimension of the corresponding stretched velocity
vector $\tilde{V}$ so that $[\tilde{X}]=[\tilde{x}-\tilde{V}\tilde{t}]$. As
a consequence the turbulent concentration density has finally the form of a
heavy tailed stretched Gaussian (for definiteness, we consider only the one
dimensional transport problem) 
\begin{equation}
C(x,t)=t^{-\frac{\alpha(\epsilon) }{2}}G_{\beta(\epsilon) }(\frac{x-Vt}{t^{\alpha(\epsilon) /2\beta(\epsilon) }})
\label{ng}
\end{equation}%
where the stretched Gaussian density is defined by \cite{klafter,chen}, $%
G_{\beta }(u)=\frac{1}{(4\pi \nu )^{\frac{1}{2}}}e^{-\frac{u^{2\beta }}{4\nu 
}}$. Using the substitution $u=\frac{x-Vt}{t^{\alpha /2\beta }}$, we now
compute the mean and variance of the turbulent advected-diffusive transport
in the form $<x>=\int_{-\infty }^{\infty }xC(x,t)dx=Vt^{1+\eta(\epsilon) }$ and $%
\sigma ^{2}(t)=\int_{-\infty }^{\infty }(x-Vt)^{2}C(x,t)dx=(4\kappa
)^{1/\beta(\epsilon) }t^{\gamma(\epsilon) }$, 
where $\eta(\epsilon) =\frac{\alpha(\epsilon) -\alpha(\epsilon) \beta(\epsilon) }{2\beta(\epsilon) }$
and $\gamma(\epsilon) =\frac{3\alpha(\epsilon) -\alpha(\epsilon) \beta(\epsilon) }{2\beta(\epsilon) }$. We note that the
normal Gaussian statistics $\eta =0,\ \gamma =1$ characterizes the transport
in a laminar flow. For turbulent transport one expects, on the other hand,
anomalous non-Gaussian statistics. We have sub-advective and sub-diffusive
behaviour for $\beta >1$ and $\alpha <2\beta (3-\beta )^{-1}$ (local dependence 
suppressed for simplicity of notation). The transport
is super-advective and super-diffusive when $0<\beta <1$ and $\alpha >2\beta
(3-\beta )^{-1}$. Incidentally, we note that the above anomalous scaling
of the mean square displacement was also obtained by Chen et al \cite{chen}
in the context of a fractal derivative diffusion problem. However, the
definition of fractal derivative appears to be
introduced in an ad hoc manner. The present approach might be considered to
be a more rigorous independent derivation of the said scaling law. 

The (multifractal) scaling exponents $\alpha $ and $\beta $
are free parameters to be fixed by matching the theoretical statistics $\eta 
$ and $\gamma $ with experimental data of a specific transport problem. The
extrapolated values of $\alpha $ and $\beta $, in turn, should be important
in distinguishing various physical characteristics of the turbulent flow and
the associated transport problem. The actual problem of comparing 
the above analytical results with real data will be considered separately. We remark, 
however, that estimating scaling exponent $\alpha(\epsilon)$, for instance, would require evaluation 
of nonlinear jump increments of the form $\alpha(\epsilon)\approx \frac{\log \eta/\tilde \eta}{\log \eta/\epsilon}$ 
(c.f. Sec. 2.2) for linear increments $\eta=\Delta t$ and $\tilde \eta=\tilde {\Delta t}$ relative to the specified scale $\epsilon$ 
and satisfying $0<\tilde \eta<\epsilon<\eta$ from a time series data. It is expected that a log-log plot of the time series data from a turbulent flow would give a nontrivial slope $\alpha\neq 1$ for each choice of the scale $\epsilon$. This fact might be interpreted as an observational validation of the duality principle advocated in this paper.

\section{Concluding Remarks}

A new mechanism of instabilities and turbulence is discussed in the context
of the kinetic Vlasov-Maxwell theory of plasma flow. The mechanism is based
on a dynamic realization of extra kinetic energy influx from an asymptotic
boundary layer region into the fluid flow. The production of extra kinetic
energy is manifested in the form of multifractal measures indicating
formation of multifractal granular structures in an unstable turbulent flow.
The mathematical formalism that supports this copious production of multifractal 
structures is presented in detail.
We illustrate the onset of  new instability in the context of the warm
plasma Langmuir wave dispersion relation. As a second example we show how
the anomalous transport and a stretched Gaussian probability density
function are derived from the turbulent advection-diffusion equation for a
plasma flow in the present scenario. The comparison of the results presented 
here with experimental data will be considered elsewhere.

\section*{Acknowlegments}
Authors thank the referee for  insightful and constructive  comments to improve the clarity and readability of the paper. Thanks are also due to A. Palit for help in drawing the figures.

\end{document}